\documentclass[10pt,conference]{IEEEtran}
\usepackage{amsmath,epsfig,amssymb,mathrsfs,psfrag,epsf,enumerate,bm}

\def\H{{\mathbf H}}
\def\HH{\widehat{{\mathbf H}}}

\def\s{{\mathbf x}}     

\usepackage{epsfig}
\usepackage{mathrsfs}
\usepackage{stmaryrd}
\usepackage{amsmath,amstext,amsfonts,amssymb}
\usepackage{epsfig,float}
\usepackage{graphicx}
\usepackage{cite}
\usepackage{psfrag}
\usepackage{subfigure}
\usepackage{url}
\usepackage{shadow,color,pifont,times,rotate}
\DeclareMathAlphabet{\mathpzc}{OT1}{pzc}{m}{it}

\newcommand{\mb}{\mathbf}

\newcommand{\mc}{\mathcal}

\newtheorem{theorem}{Theorem}[section]
\newtheorem{lemma}[theorem]{Lemma}

\title{On the Outage Capacity of a Practical Decoder Using Channel Estimation Accuracy}
\author{\authorblockN{Pablo  Piantanida}
\authorblockA{CNRS/LSS - Sup\'elec\\
Gif-sur-Yvette 91192, France \\
Email: piantanida@lss.supelec.fr}
   \and
\authorblockN{Sajad Sadough}
\authorblockA{ENSTA - CNRS/LSS\\
32 bd Victor 75015 Paris, France\\
Email: sajad.sadough@ensta.fr}
\and
\authorblockN{Pierre Duhamel}
\authorblockA{CNRS/LSS - Sup\'elec\\
Gif-sur-Yvette 91192, France\\
Email: pierre.duhamel@lss.supelec.fr}
}

\begin{document}
\maketitle

\begin{abstract}
The optimal decoder achieving the outage capacity under imperfect
channel estimation is investigated. First, by searching into the family
 of nearest neighbor decoders, which can be easily implemented on most practical
 coded modulation systems, we derive a decoding metric that minimizes the average of
 the transmission error probability over all channel estimation errors. This metric,
  for arbitrary memoryless channels, achieves the capacity of a composite (more noisy) channel.
  Next, according to the notion of estimation-induced outage capacity (EIO capacity)
  introduced in our previous work, we characterize maximal achievable information
  rates associated to the proposed decoder. The performance of the proposed decoding
   metric over uncorrelated Rayleigh fading MIMO channels is compared to both the classical
    mismatched maximum-likelihood (ML) decoder and the theoretical limits given by the EIO
    capacity (i.e. the best decoder in presence of channel estimation errors). Numerical
    results show that the derived metric provides significant gains, in terms of achievable
     information rates and bit error rate (BER), in a bit interleaved coded modulation (BICM)
      framework, without introducing any additional decoding complexity.
\end{abstract}
\vspace{-1mm}
\section{Introduction}


Consider practical wireless communication systems, where each receiver disposes only of noisy channel estimates that may in some circumstances be poor estimates, and these estimates are not available at the transmitter. This constraint constitutes a practical concern for the design of such communication systems that, in spite of their knowledge limitations, have to ensure communications with a prescribed quality of service (QoS). This QoS requires to guarantee communications with a given target information rate and small error probability, no matter which degree of accuracy estimation arises during the transmission. The described scenario addresses two important questions: (i) What are the theoretical limits of reliable transmission rates, using the best possible decoder in presence of imperfect channel state information at the receiver (CSIR) and (ii) how those limits can be achieved by using practical decoders in coded modulation systems ? Of course, these questions are strongly related to the notion of capacity that must take into account the above mentioned constraints.

Recently in \cite{isita2006}, we have addressed the first question (i) for general memoryless channels, by introducing the notion of \emph{Estimation-induced outage capacity} (EIO capacity). Basically, we consider that a specific instance of the unknown memoryless channel, with input $x \in\mathscr{X}$ and output $y\in \mathscr{Y}$, is characterized by a transition probability $W(y|x,\theta)\in\mc{W}_{\Theta}$ with an unknown channel state $\theta$, which follows i.i.d. ~$\theta\sim\psi(\theta)$; $\mc{W}_{\Theta}$ is a family of conditional pdf parameterized by the vector of parameters $\theta\in \Theta\subseteq \mathbb{C}^d$. The receiver only knows an estimate $\hat{\theta}$ and a characterization of its quality, in terms of the conditional pdf $\psi(\theta|\hat{\theta})$. A decoder using $\hat{\theta}$, instead of $\theta$, obviously might not support an information rate $R$ (even small rates might not be supported if $\hat{\theta}$ and $\theta$ are strongly different). Consequently, outages induced by channel estimation errors (CEE) will occur with a certain probability $\gamma_{_{QoS}}$.

The second question (ii) concerning the derivation of a practical decoder that, using imperfect channel estimation, can achieve information rates closed to the EIO capacity is addressed in this paper. Classically, to deal with imperfect channel state information (CSI) one sub-optimal technique, known as mismatched maximum-likehood (ML) decoding, consists in replacing the exact channel by its estimate in the decoding metric. However, this scheme is not adapted to the presence of CEE, at least for systems with small training overhead. As an alternative to this, Tarokh {\it et al.} \cite{Tarokh-1999} and Taricco and Biglieri \cite{taricco-biglieri-2005}, proposed an improved ML detection metric and applied it to a space-time coded MIMO system. This metric can be formally derived as a special case of the general framework presented here. In this paper, according to the notion of EIO capacity we derive the general expression of a decoder that minimizes the average of the transmission error probability over all CEE and consequently it achieves the capacity of a composite (more noisy) channel. Then, we evaluate this for Rayleigh fading MIMO channels and investigate maximal achievable information rates. 


\subsection{A Brief Review of Estimation-induced Outage Capacity}
\def\Mcb{M}

A message $m\in\mc{M}=\{1,\dots,\lfloor\exp(nR)\rfloor \}$ is transmitted using a pair $(\varphi,\phi)$ of mappings, where $\varphi: \mc{M}     \mapsto   \mathscr{X}^n$ is the encoder, and $\phi: \mathscr{Y}^n\times  \Theta  \mapsto   \mc{M}$ is the decoder (that utilizes $\hat{\theta}$). The random rate, which depends on the unknown channel realization $\theta$ through its probability of error, is given by $n^{-1}\log \Mcb_{\theta,\hat{\theta}}$. The maximum error probability \vspace{-.2cm}
\begin{equation}
e_{\max}(\varphi,\phi,\hat{\theta};\theta)=\max_{m\in\mc{M}}W^n\big(\{\phi(\mathbf{y}, \hat{\theta})\neq m\}\big|\varphi(m),\theta\big),\label{eq-error}\vspace{-.2cm}
\end{equation}
For a given channel estimate $\hat{\theta}$, and $0<\epsilon,\gamma_{_{QoS}}< 1$, an outage rate $R\geq 0$ is $(\epsilon,\gamma_{_{QoS}})$-achievable if for every $\delta>0$ and every sufficiently large $n$ there exists a sequence of length-$n$ block codes such that the rate satisfies\vspace{-.1cm}
\begin{equation}
\Pr\left(\Lambda_\epsilon(R,\hat{\theta}) \big | \hat{\theta} \right)=\int_{\Lambda_\epsilon(R,\hat{\theta})}\!\!\!  d\psi(\theta|\hat{\theta}) \geq 1-\gamma_{_{QoS}},\label{eq-proba}\vspace{-.1cm}
\end{equation}
where $\Lambda_\epsilon(R,\hat{\theta})=\big\{\theta\in \Delta_{\epsilon}: n^{-1}\log \Mcb_{\theta,\hat{\theta}}\,\geq\, R-\delta \big\}$, and $\Delta_{\epsilon} = \big\{\theta\in\Theta\!: \,e_{\max}(\varphi,\phi,\hat{\theta};\theta)\leq\epsilon\big\}$ is the set of all channel states allowing for reliable decoding.
This definition requires that maximum error probabilities larger than $\epsilon$ occur with probability less than $\gamma_{_{QoS}}$. The practical advantage of such definition is that for $(1-\gamma_{_{QoS}})\%$ of estimates, the transmitter and receiver strive to construct codes for ensuring the desired communication service. The  EIO capacity is then defined as the largest $(\epsilon,\gamma_{_{QoS}})$-achievable rate, for an outage probability $\gamma_{_{QoS}}$ and a given estimated $\hat{\theta}$, as\vspace{-.1cm}
\begin{equation}
C(\gamma_{_{QoS}},\hat{\theta})=\max\limits_{P\in\mathscr{P}_\Gamma(\mathscr{X})}\sup\limits_{\Lambda\subset \Theta:\,\,\Pr(\Lambda|\hat{\theta})\geq 1-\gamma_{_{QoS}}}\!\!\!\!\!\!\inf\limits_{\theta\in\Lambda}I\big(P,W(\cdot|\cdot,\theta)\big),\label{eq-max}\vspace{-.1cm}
\end{equation}
where $I(\cdot)$ denotes the mutual information of the channel $W(y|x,\theta)$ and $\mathscr{P}_\Gamma(\mathscr{X})$ is the set of input distributions not depending on $\hat{\theta}$. The theoretical decoder achieving the capacity \eqref{eq-max}, based on the well-known method of typical sequences, cannot be implemented on practical communication systems. Indeed in \cite{spawc06}, the achievable rates obtained with the mismatched ML decoding have been showed to be largely smaller compared to the EIO capacity.

\subsection{A Practical Decoder Using Channel Estimation Accuracy}

We now consider the problem of deriving a practical decoder that achieves the capacity \eqref{eq-max}. Assume that we limit the searching of decoding functions $\phi$ to the class of additive decoding metrics, which can be implemented on realistic systems. This means that for a given channel output $\mb{y}=(y_1,\dots,y_n)$, we set the decoding function
\begin{equation}
\phi_{\mc{D}}(\mb{y},\hat{\theta})=\arg\min\limits_{m\in \mc{M}}\mc{D}^n\big(\varphi(m),\mb{y}|\hat{\theta}\big), \label{decoder-metric}\vspace{-1mm}
\end{equation}
where $\mc{D}^n\big(\mb{x},\mb{y}|\hat{\theta}\big)=\frac{1}{n}\sum_{i=1}^n \mc{D}\big(x_i,y_i|\hat{\theta}\big)$ and $\mc{D}: \mathscr{X}\times \mathscr{Y}\times \Theta  \mapsto \mathbb{R}_{\geq 0}$ is an arbitrary per-letter additive metric.  Consequently, the maximization in \eqref{eq-max} is actually equivalent to maximizing over all decoding  metrics $\mc{D}$. However, we note that this restriction does not necessarily lead to an optimal decoder achieving the capacity.

In order to find the optimal decoding metric $\mc{D}$ maximizing the outage rates, for a given probability $\gamma_{_{QoS}}$ and estimate $\hat{\theta}$, it is necessary to look at the intrinsic properties of the capacity definition. Observe that the size of the set of all channel states allowing for reliable decoding $\Delta_{\epsilon}$ is determined by the decoding function $\phi$ chosen and the maximal achievable rate $R$, constrained to the outage probability \eqref{eq-proba}, is then limited by this size. Thus, for a given decoder $\phi$, there exists an optimal set $\Lambda^*_\epsilon \subseteq \Delta_{\epsilon}$ of channel states with conditional probability larger than $1-\gamma_{_{QoS}}$, providing the largest achievable rate, which follows as the minimal instantaneous rate for the worst $\theta \in\Lambda^*_\epsilon$. The optimal set $\Lambda^*_\epsilon$ is equal to the set $\Lambda^*$ maximizing the expression \eqref{eq-max}. Hence, an optimal decoding metric must guarantee minimum error probability \eqref{eq-error} for every $\theta \in\Lambda^*$. Then, the computation of such metric becomes very difficult, since the maximization in \eqref{eq-max} by using $\phi_\mc{D}$ is not really an explicit function of $\mc{D}$.

Instead of trying to find an optimal decoding metric minimizing the transmission error probability \eqref{eq-error} for every $\theta \in\Lambda^*$, we propose to look at the decoding metric minimizing the average of this error probability over all CEE. This means,\vspace{-1mm}
\begin{equation}
\mc{D}_\mc{M}=\arg\min\limits_{\mc{D}} \int_\Theta e_{\max}^{(n)}(\varphi,\phi_{\mc{D}},\hat{\theta};\theta) d\psi(\theta|\hat{\theta}),\label{eq_average_error}\vspace{-2mm}
\end{equation}
where $e_{\max}^{(n)}$ follows by replacing \eqref{decoder-metric} in \eqref{eq-error}. Actually,  for $n$ sufficiently large, this optimization problem can be resolved by setting \vspace{-2mm}
\begin{eqnarray}
\mc{D}_\mc{M}(x,y|\hat{\theta})=-\log \widetilde{W}(y|x,\hat{\theta}),\label{metric-definition}\vspace{-2mm}
\end{eqnarray}
$\widetilde{W}(y|x,\hat{\theta})=\int_\Theta W(y|x,\theta) d\psi(\theta|\hat{\theta})$ is the channel resulting from the average of the unknown channel over all CEE, given the estimate $\hat{\theta}$. Here we do not go into the details of how the optimal metric \eqref{metric-definition} minimizes \eqref{eq_average_error}. Basically, the average of the transmission error probability leads to the composite (more noisy) channel $\widetilde{W}(y|x,\hat{\theta})$, and then we take the logarithm of this composite channel to obtain its ML decoder.

In the remainder of this paper, we evaluate the derived decoding metric \eqref{metric-definition} for uncorrelated Rayleigh fading MIMO channels and use it in a bit interleaved coded modulation (BICM) receiver (section II). Then, we compute the achievable rates according to the considered notion of the EIO capacity (section III). In section IV, we illustrate via numerical simulations the performance of the improved decoder and compare it to the mismatched ML decoder.
\section{Channel Model, Decoding with imperfect CSIR and Receiver Processing}

We use upper case and lower case boldface letter for matrix and vectors, respectively; $\|\cdot\|_F$ denotes the Frobenius norm, $\textrm{diag}(\mb{x})$ denotes a diagonal matrix with elements $\mb{x}$, $\textrm{diag}(\H)$ denotes the vector corresponding to the diagonal elements of the matrix $\H$ and $(\cdot)^{\dag}$ the Hermitian transposition.

\subsection{MIMO Channel model}

Consider a single-user memoryless Fading MIMO channel with $M_T$ transmitter and $M_R$ receiver antennas. The discrete-time channel at time $t$ is modeled by\vspace{-2mm}
\begin{equation}
\mb{y}(t)=\H(t)\mb{x}(t)+\mb{z}(t),\vspace{-1mm}\label{channel-definition}
\end{equation}
where $\mb{x}(t) \in \mathbb{C}^{M_T\times 1}$ is the vector of transmitter symbols and $\mb{y}(t)\in \mathbb{C}^{M_R\times 1}$ is the vector of received symbols; $\theta=\H(t)\in \mathbb{C}^{M_R\times M_T}$ is the complex random matrix whose entries are i.i.d. zero-mean circularly symmetric complex Gaussian (ZMCSCG) random variables $C\mc{N}(0,\sigma^2_{h})$. The channel is a complex normal distributed matrix $\H(t)\!\sim \psi(\theta)=C\mc{N}\big(0, \mathbb{I}_{M_T}\otimes \Sigma_{H}\big)$,  with $\Sigma_{H}=\sigma_{h}^2\mathbb{I}_{M_R}$. The noise vector $\mb{z}(t)\in \mathbb{C}^{M_R\times 1}$ consists in ZMCSCG random vector with covariance matrix $\Sigma_{0}=\sigma_{Z}^2\mathbb{I}_{M_R}$. This leads to a channel $W(\mb{y}|\mb{x},\H)=C\mc{N}\big(\H\mb{x},\Sigma_{0}\big)$. The input symbols are constrained to satisfy $tr\big(\mathbb{E}_\mb{x}(\mb{x}(t) \mb{x}(t)^\dag)\big)\leq \bar{P}$.

\emph{Channel estimation:} We assume that the transmitter, before sending the data $\mb{x}$, can teach the channel to the receiver by sending a training sequence of $N$ vectors $\mb{X}_T=(\mb{x}_{T,1},\dots,\mb{x}_{T,N})$. We assume that the coherence time of the channel is much longer than the training time and the average energy of the training symbols is $P_T=\frac{1}{N M_T}tr\big(\mb{X}_T \mb{X}_T^\dag\big)$.  This sequence is affected by the channel matrix $\mb{H}$, allowing the receiver to perform ML estimation of $\mb{H}$ from the observed signals $\mb{Y}_{T}=\mb{H}\mb{X}_T+\mb{Z}_{T}$ and $\mb{X}_T$. This yields to $\hat{\theta}=\widehat{\mb{H}}=\mb{H}+\mc{E}$, where $\mc{E}$  denotes the estimation error matrix yielding to a white error matrix $\Sigma_{\mc{E}}=\sigma_{\mc{E}}^2\mathbb{I}_{M_R}$ and $\sigma_{\mc{E}}^2=\textrm{SNR}^{-1}_{T}$ with $\textrm{SNR}_{T}=\frac{NP_T}{\sigma^2_{Z}}$, when the training sequences are orthogonal.

\emph{Mismatched ML decoder:}  The classical mismatched ML decoder consists of the likelihood function of the channel using the channel estimate $\HH$, $\mc{D}_{\textrm{ML}}\big( \mb{x}, \mb{y}|  \HH \big)=-\log W(\mb{y}|\mb{x},\HH)$. This leads to the following Euclidean distance\vspace{-1mm}
\begin{equation}
\mc{D}_{\textrm{ML}}\big( \mb{x}, \mb{y}|  \HH \big)=\|\mb{y}-\widehat{\mb{H}}\mb{x}\|^2+\textrm{const}.\label{mismached_ML}\vspace{-1mm}
\end{equation}

\subsection{Metric computation}

We now evaluate the general metric expression of \eqref{metric-definition} in the case of a MIMO channel model \eqref{channel-definition}. To this end, we first derive the pdf $\psi(\theta|\hat{\theta})$, which can be obtained by using the likelihood function, the pdf $W(\mb{y}|\mb{x},\H)$, and $\psi(\theta)$. Then, by averaging the channel $W(\mb{y}|\mb{x},\H)$ over all CEE and after some algebra, we obtain  the channel $\widetilde{W}(\mb{y}|\mb{x},\HH )=C\mc{N}\big( \delta\HH \mb{x}, \Sigma_{0} + \delta\Sigma_{\mc{E}}   \| \mb{x} \|^2     \big)$
where $\delta=\frac{\textrm{SNR}_{T}\sigma^2_{h}}{\textrm{SNR}_{T}\sigma^2_{h}+1}$. Finally, from \eqref{metric-definition} the optimal decoding metric for this channel is reduced to\vspace{-1mm}
\begin{equation}
\mc{D}_\mc{M}\big( \mb{x}, \mb{y}|  \HH \big)=M_R \log ( \sigma^2_Z+\delta \sigma_{\mc{E}}^2 \| \mb{x}\|^2)+ \frac{\| \mb{y}- \delta \HH  \mb{x}  \|^2}{\sigma^2_Z+\delta \sigma_{\mc{E}}^2 \| \mb{x}\|^2},\label{final_metric_MIMO}\vspace{-1mm}
\end{equation}
and this metric coincides with that proposed for space-time decoding, from independent results in \cite{Tarokh-1999}  and \cite{taricco-biglieri-2005}.

\subsection{Receiver structure}
\label{sec:bicmRX}

The problem of decoding MIMO-BICM has been addressed in \cite{boutros00} under the assumption of perfect CSIR. Here we consider the same problem with CEE, for which we use the metric \eqref{final_metric_MIMO} to the iterative decoding process of BICM. Basically, the receiver consists of the combination of two sub-blocks operating successively. The first sub-block, referred to as soft symbol to bit MIMO demapper, produces bit metrics (probabilities) from the input symbols and the second one is a soft-input soft-output (SISO) trellis decoder. Each sub-block can take advantage of the a posteriori (APP) provided by the other sub-block as an additive information. Here, SISO decoding is performed using the well known forward-backward algorithm \cite{bcjr}. We recall the formulation of the soft MIMO detector.

Suppose first the case where the channel matrix $\H_k$ is perfectly known at the receiver. The MIMO demapper provides at its output the extrinsic probabilities on coded and interleaved bits $\mb{d}$. Let $d_{k,i}$, $i = 1,\dots,BM_T$, be the interleaved bits corresponding to the $k$-th compound symbol $\mb{x}_k \in Q$ where the cardinality of $Q$ is $\|Q\| = 2^{BM_T}$. The extrinsic probability $P_{\rm dem}(d_{k,j})$ of the bit $d_{k,j}$ (bit metrics) at the MIMO demapper output is calculated as \vspace{-2mm}
\begin{equation}\label{eq:mimoRx1}
P_{\rm dem}(d_{k,j} = 1) = K \sum_{\stackrel{\mb{x}_k\in Q} {d_{j}=1}}  \prod_{\stackrel{i=1}{i\neq j}}^{B M_T} P_{\rm dec}(d_{i})\exp\big[-\mc{D}(\mb{x}_k,\mb{y}_k| \H_k) \big],\vspace{-2mm}
\end{equation}
where $K$ is the normalization factor satisfying $P_{\rm dem}(d_{k,j} = 1) + P_{\rm dem}(d_{k,j} = 0) = 1$ and $P_{\rm dec}(d_{k,j})$ is the {\it prior} information on bit $d_{k,j}$, coming from the SISO decoder. The summation in \eqref{eq:mimoRx1} is taken over the product of the channel likelihood given a compound symbol $\mb{x}_k$, and the {\it a priori} probability on this symbol (the term $\prod P_{\rm dec}$) feedback from the SISO decoder at the previous iteration. Concerning this latter term, the {\it a priori} probability of the bit $d_{k,j}$ itself has been excluded, so as to let the exchange of extrinsic information between the channel decoder and the MIMO demapper.  At the first iteration we set $P_{\rm dec}(d_{k,i})=1/2$. Notice that by replacing the unknown channel involved in \eqref{eq:mimoRx1} by its estimate $\HH_k$, we obtain the mismatched ML decoder of MIMO-BICM. Instead of this, we introduce in \eqref{eq:mimoRx1} the demaping rule $\mc{D}_{\mc{M}}$ \eqref{final_metric_MIMO}, which is adapted to the CEE.

\section{Computation of Achievable Information Rates}

We now derive the achievable rates $C_{\mathcal{D}}$ associated to a receiver using the decoding rule \eqref{decoder-metric}, based on the derived metric \eqref{final_metric_MIMO}. This is done by using the following Theorem \cite{merhav-1994}, for the considered channels $W(\mathbf{y}|\s,\H)=C\mc{N}\big(\H \mb{x},\Sigma_{0}\big)$.
\newtheorem{theo}{Theorem}
\begin{theorem}\label{theo_merhav}
For any pair of matrices $({\H,\HH})$, the maximal achievable rate associated to a receiver using a metric $\mc{D}(\s,\mb{y}|\widehat{\H})$ is given by \vspace{-1mm}
\begin{equation}
\label{eq:outage}
C_{\mc{D}}(\H,\HH)=\sup_{P_X\in \mathscr{P}_\Gamma(\mathscr{X})}  \inf_{V_{Y|X} \in \mc{V}(\H,\widehat{\H})} I( P_X,V_{Y|X}),\vspace{-1mm}
\end{equation}
where the mutual information functional $I( P_X,V_{Y|X})=$\vspace{-1mm}
\begin {equation}
\label{eq:I}
\int\!\!\!\!  \int \!  \log_2 \frac{  V_{Y|X} (\mathbf{y}|\mathbf{x},\Upsilon)} {\int   V_{Y|X} (\mathbf{y}|\mathbf{x}^\prime,\Upsilon) dP_X(\mathbf{x}^\prime)}dP_X(\mathbf{x}) dV_{Y|X} (\mathbf{y}|\mathbf{x},\Upsilon),
\end{equation}
and $\mathcal{V}(\H,\widehat{\H})$ denotes the set of test channels, i.e., all possibles uncorrelated MIMO channels $V_{Y|X} (\mathbf{y}|\mathbf{x},\Upsilon)=C\mathcal{N}(\Upsilon \mb{x},\mb{\Sigma})$, verifying that\vspace{-2mm}
\begin{align*}
&(c_1): tr \big( \mathbb{E}_P  \, \big\{\mathbb{E}_{V} \{\mb{Y}\mb{Y}^\dag \} \big\} \big) = tr \big(   \mathbb{E}_P \, \big\{ \mathbb{E}_{W}\{ \mb{Y}\mb{Y}^\dag \}\big\} \big),
\\  \label{eq:ineqconstr1}
&(c_2): \mathbb{E}_P  \, \Big\{\mathbb{E}_{V} \big\{\mathcal{D}(\s,\mb{y}|\widehat{\H})\big\} \Big\} \leq \mathbb{E}_P \, \Big\{ \mathbb{E}_{W} \big\{\mathcal{D}(\s,\mb{y}|\widehat{\H})\big\}\Big\}.
\end{align*}
\end{theorem}\vspace{1mm}

\emph{Computation of achievable rates:} In order to solve the constrained minimization problem in Theorem \eqref{theo_merhav} for our metric $\mc{D}=\mc{D}_\mc{M}$ (expression \eqref{final_metric_MIMO}), we must find the channel $\Upsilon\in \mathbb{C}^{M_R\times M_T}$ and the covariance matrix $\mb{\Sigma}=\mathbb{I}_{M_R} \sigma^2$ defining the test channel $V_{Y|X} (\mathbf{y}|\mathbf{x},\Upsilon)$ that minimizes the mutual information \eqref{eq:I}. On the other hand, through this paper we assume that the transmitter does not dispose of the channel estimates, and consequently no power control is possible. Thus, we choose the sub-optimal input distribution $P_X=C\mathcal{N}(\mb{0},\mb{\Sigma_P})$ with $\mb{\Sigma_P}= \mathbb{I}_{M_T} \bar{P}$. We first compute the constraint set $\mc{V}(\H,\widehat{\H})$, given by $(c_1)$ and $(c_2)$, and then we factorize the matrix $\H$ to solve the minimization problem. Before this, we state the following result that, due to lack of space, we do not include it in this paper.

\begin{lemma}\label{elemma}
Let $\mb{A}\in \mathbb{C}^{M_R\times M_T}$ be an arbitrary matrix and $\mb{X}$ be a random vector with pdf $C\mathcal{N}(0,\mb{\Sigma_P})$. For every real positive constants $K_1,K_2>0$, the following equality holds
$$
\mathbb{E}_{\mb{X}}\!\! \left[ \frac{\| \mb{A} \mb{X}  \|^2 + K_1}{\| \mb{X}\|^2+K_2} \right]=\frac{\|\mb{A}\|_F^2}{n+1} + \left ( \frac{K_1}{K_2}-\frac{\|\mb{A}\|_F^2}{n+1} \right) \left(\frac{K_2}{\bar{P}}\right)^{n+1}\times
$$
$ \displaystyle{\exp  \left( \frac{K_2}{\bar{P}} \right) \Gamma\left(-n,K_2/\bar{P}\right)}$, $n=M_T-1$ with $n\in \mathbb{N}_+$ and $\Gamma(-n,t)=\frac{(-1)^n}{n !} \Big[\Gamma(0,t)-\exp(-t)\sum\limits_{i=0}^{n-1}(-1)^i \frac{i !}{t^{i+1}} \Big]$ and $\Gamma(0,t)$ denotes the exponential integral function.
\end{lemma}
Then, by using Lemma \ref{elemma} and some algebra, we have that
\begin{eqnarray}
&\mathrm{(c_1)}:&  tr \big(\Upsilon\mb{\Sigma_P} \Upsilon^\dag+ \mb{\Sigma} \big) = tr \big(\H \mb{\Sigma_P} \H^\dag+ \mb{\Sigma_0} \big), \label{c1}\\
&\mathrm{(c_2)}:& \| \Upsilon +a_\mc{M} \widehat{\H} \|^2_F\leq  \| \H +a_\mc{M} \widehat{\H} \|^2_F + \textrm{C},\label{c2}
\end{eqnarray}
where $a_\mc{M}=\delta(\delta\sigma^2_{\mathcal{E}} \bar{P} - \lambda_n \sigma_Z^2)\big[M_T\delta\sigma^2_{\mc{E}}\lambda_n \bar{P}+\lambda_n \sigma_Z^2- \delta\sigma^2_{\mathcal{E}} \bar{P} \big]^{-1}$, $\textrm{C}= M_T \lambda_n \big[\|\H\|^2_F-\|  \Upsilon\|^2_F + \bar{P}^{-1}\big(tr(\mb{\Sigma_0})-tr(\mb{\Sigma})\big)\big] \big[1-\frac{\sigma_Z^2}{\delta \bar{P}\sigma^2_{\mathcal{E}}}\lambda_n-M_T \lambda_n  \big]^{-1}$ and $\lambda_n=\left(\frac{ \sigma_Z^2}{\delta \bar{P} \sigma^2_{\mathcal{E}} }\right)^n \exp\left(\frac{ \sigma_Z^2}{\delta \bar{P} \sigma^2_{\mathcal{E}} }\right)\Gamma\left(-n,\frac{ \sigma_Z^2}{\delta \bar{P}\sigma^2_{\mathcal{E}} }\right)$.\vspace{1mm}


From expression \eqref{c2} and computing the mutual information, the minimization in \eqref{eq:outage} writes as $C_{\mc{M}}^\textrm{MIMO}(\H,\widehat{\H})=$
\begin{equation}
\left \{ \begin{array}{ll} \min\limits_{\Upsilon} \,\,\,\,\,\, \log_2 \textrm{det}\left(\mathbb{I}_{M_R}+ \Upsilon \mb{\Sigma_P} \Upsilon^\dag \mb{\Sigma}^{-1}\right), \\ \textrm{subject to} \,\,\,\,\,\,  \| \Upsilon +a_\mc{M} \widehat{\H} \|^2_F\leq  \| \H +a_\mc{M} \widehat{\H} \|^2_F + \textrm{C}, \end{array}\right.\label{eq_optb2}
\end{equation}
where $\mb{\Sigma}$ must be chosen such that $tr \big(\Upsilon\mb{\Sigma_P} \Upsilon^\dag+ \mb{\Sigma} \big) = tr \big(\H \mb{\Sigma_P} \H^\dag+ \mb{\Sigma_0} \big)$. In order to obtain an alternative expression of \eqref{eq_optb2}, simpler and more tractable, we consider the following decomposition of the matrix $\H=\mathbf{U}\, \textrm{diag}(\underline{\lambda}) \mathbf{V}^\dag$ with $\underline{\lambda}=(\lambda_1,\dots,\lambda_{M_R})^T$. Let $\textrm{diag}(\underline{\mu})$ be a diagonal matrix such that $\textrm{diag}(\underline{\mu})=\mathbf{U}^\dag \Upsilon \mathbf{V}$, whose diagonal values are given by the vector $\underline{\mu}=(\mu_1,\dots,\mu_{M_R})^T$. We define $\widetilde{\H}^\dag=\mathbf{V}^\dag \widehat{\H}^\dag \mathbf{U}$, the vector $\mb{\tilde{h}}^\dag=\textrm{diag}(\widetilde{\H}^\dag)^T$ resulting of its diagonal and let $b_\mc{M}= \| \H + a_\mc{M} \widehat{\H}\|_F^2 - a_\mc{M}^2 (\| \widetilde{\H}\|_F^2-\|  \mb{\tilde{h}}\|^2)$. Using the above definitions and some algebra, the optimization \eqref{eq_optb2} writes
\begin{equation}
C_{\mc{M}}^\textrm{MIMO}(\H,\widehat{\H})=\left \{ \begin{array}{ll} \min\limits_{\underline{\mu}} \,\,\,\,\,\,  \sum\limits_{i=1}^{M_R}\log_2 \left(1+\displaystyle{\frac{\bar{P}|\mu_i|^2}{\sigma^2( \underline{\mu})}}\right), \\ \textrm{subject to} \,\,\,\,\,\,  \| \underline{\mu} +a_\mc{M}  \mb{\tilde{h}}  \|^2\leq b_\mc{M}, \end{array}\right.\label{final_opt}
\end{equation}
with $\sigma^2( \underline{\mu})=\frac{\bar{P}}{M_R}(\|\underline{\lambda} \|^2-\|\underline{\mu} \|^2)+\sigma_Z^2$. The constraint set in the minimization \eqref{final_opt}, which corresponds to the set of vectors $\{\underline{\mu}\in \mathbb{C}^{M_T\times 1} :\,   \| \underline{\mu} +a_\mc{M}  \mb{\tilde{h}}  \|^2\leq b_\mc{M}\}$, is a closed convex polyhedral set. Thus, the infimun in \eqref{final_opt} is attainable at the extremal of the set given by the equality (cf. \cite{convex-book}). On the other hand, for every vector $\underline{\mu}$ such that $\|\underline{\mu} \|^2 \leq  \|\underline{\lambda}\|^2$, we observe that the expression \eqref{final_opt} is a monotone increasing function of the square norm of $\underline{\mu}$. As a consequence, it is sufficient to find the optimal vector $ \underline{\mu}^{\textrm{opt}}_\mc{M}$ minimizing the square norm over the constraint set. This becomes a classical convex minimization problem that can be easily solved by using Lagrange multipliers. The corresponding achievable rates are then given by \vspace{-1mm}
\begin{equation}
C_{\mc{M}}^\textrm{MIMO}(\H,\widehat{\H})=\log_2 \textrm{det}\left(\mathbb{I}_{M_R}+ \Upsilon_{\textrm{opt}} \mb{\Sigma_P}\Upsilon_{\textrm{opt}}^\dag \sigma^{-2}( \underline{\mu}^{\textrm{opt}}_\mc{M})   \right),\label{acievable_rates_MIMO}
\end{equation}
where the optimal solution $\Upsilon_{\textrm{opt}}=\mathbf{U}\, \textrm{diag}(\underline{\mu}^{\textrm{opt}}_\mc{M}) \mathbf{V}^\dag$ with
\begin{equation}
\underline{\mu}^{\textrm{opt}}_\mc{M}= \left\{ \begin{array}{ll} \left(\frac{\sqrt{b_\mc{M}}}{\|\mb{\tilde{h}}\|}-|a_\mc{M}|\right)\mb{\widetilde{h}} & \,\, \textrm{if $b_\mc{M}\geq 0$,} \\
  \underline{0} & \,\, \textrm{otherwise}, \end{array}  \right. \label{solution_mu}
\end{equation}

For any pair of matrices $(\H,\HH)$, the expression \eqref{acievable_rates_MIMO} provides the instantaneous achievable rates associated to a receiver using the decoding rule \eqref{decoder-metric}, based on the derived metric \eqref{final_metric_MIMO}.  Whereas, the outage rates  $R \geq 0$ must be computed by using the associated outage probability $P_{\mathcal{D}}^{\mathrm{out}}$. This outage probability is defined by \eqref{eq-proba}, where here $\Lambda_{\mathcal{D}}(R,\widehat{\H})=\big\{\H:\, C_{\mathcal{D}}(\H,\widehat{\H})<R\big\}$. Therefore, the maximization of the outage rate \eqref{eq-max}, for an outage probability $\gamma_{_{QoS}}$, is given by
\begin{equation}
\label{capout}
C_{\mathcal{D}}(\gamma_{_{QoS}},\widehat{\H})=\sup\big\{R \geq 0: P_{\mathcal{D}}^{\mathrm{out}}(R,\widehat{\H})\leq \gamma_{_{QoS}}\big\}.
\end{equation}

Before conclude this section, following the same steps as above, we can compute the achievable rates associated to the mismatched ML decoder \eqref{mismached_ML}. These are given by replacing in expression \eqref{acievable_rates_MIMO}  the solution vector \vspace{-2mm}
\begin{equation}
\label{mismached_ML_capa}
\underline{\mu}^{\textrm{opt}}_\textrm{ML}=\frac{ \mathbb{R}e\{tr(\Lambda^\dag \tilde{\mb{h}}) \}}{\|\tilde{\mb{h}}\|^2}\tilde{\mb{h}}.
\end{equation}

\section{Numerical Results}
\label{sec:simul}

In this section we provide numerical results to analyze the performance of a receiver using the derived metric \eqref{final_metric_MIMO}, over uncorrelated block fading MIMO channels. The performances are measured in terms of BER and achievable outage rates. The binary information data are encoded by a rate $1/2$ non-recursive non-systematic convolutional channel code with constraint length 3 defined in octal form by (5,7). Throughout the simulations, each frame is assumed to consists of 100 MIMO symbols belonging to a 16-QAM constellation with Gray labeling. The interleaver is a random one operating over the entire frame with size $100\! \cdot\! M_T\! \cdot\! \log_2(B)$ bits. Although longer interleavers are expected to yield somewhat improved performance, this length was choosed because of latency requirements. For each transmitted frame, a different realization of the channel has been drawn and remains constant during the whole frame. Besides, it is assumed that the average pilot symbol energy is equal to the average data symbol energy.

\begin{figure}[!htb]
\vspace{-3mm}
\centering
\includegraphics[width=0.5\textwidth,height=0.25\textheight]{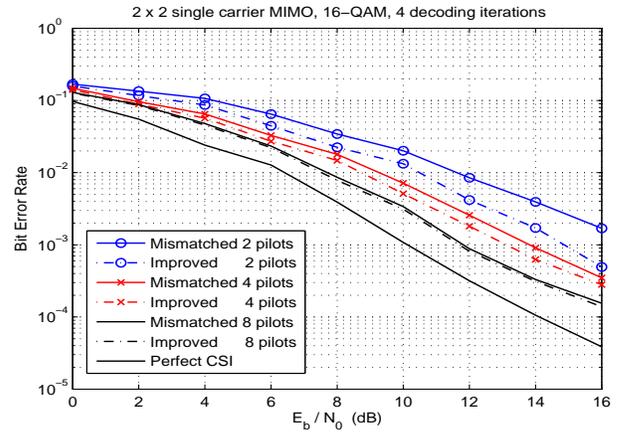}
\caption{BER performance of $2\times2$ MIMO of the proposed decoder over Rayleigh fading channel for various training sequence lengths.} \label{ber22}
\end{figure}

Fig. \ref{ber22}. shows the increase in required $E_b/N_0$ caused by the CEE for a $2\times2$ MIMO channel in the case of mismatched ML decoding. We insert $N=2,4$ or $8$ pilot symbols per frame for CSIR acquisition. At ${\rm BER} = 10^{-3}$ and $N=2$, we observe about $2$ dB of SNR gain by using the improved  decoder. We also notice that the performance loss of the mismatched receiver with respect to the derived receiver becomes insignificant for $N\geq8$. This can be explained from the expression of the metric \eqref{final_metric_MIMO}, where we note that by increasing the number of pilot symbols, this expression tends to the classical Euclidean distance metric (see equation \eqref{mismached_ML}). This clearly shows that the investigated decoder outperforms the mismatched decoder.

\begin{figure}[!htb]
\vspace{-4mm}
\centering
\includegraphics[width=0.5\textwidth,height=0.25\textheight]{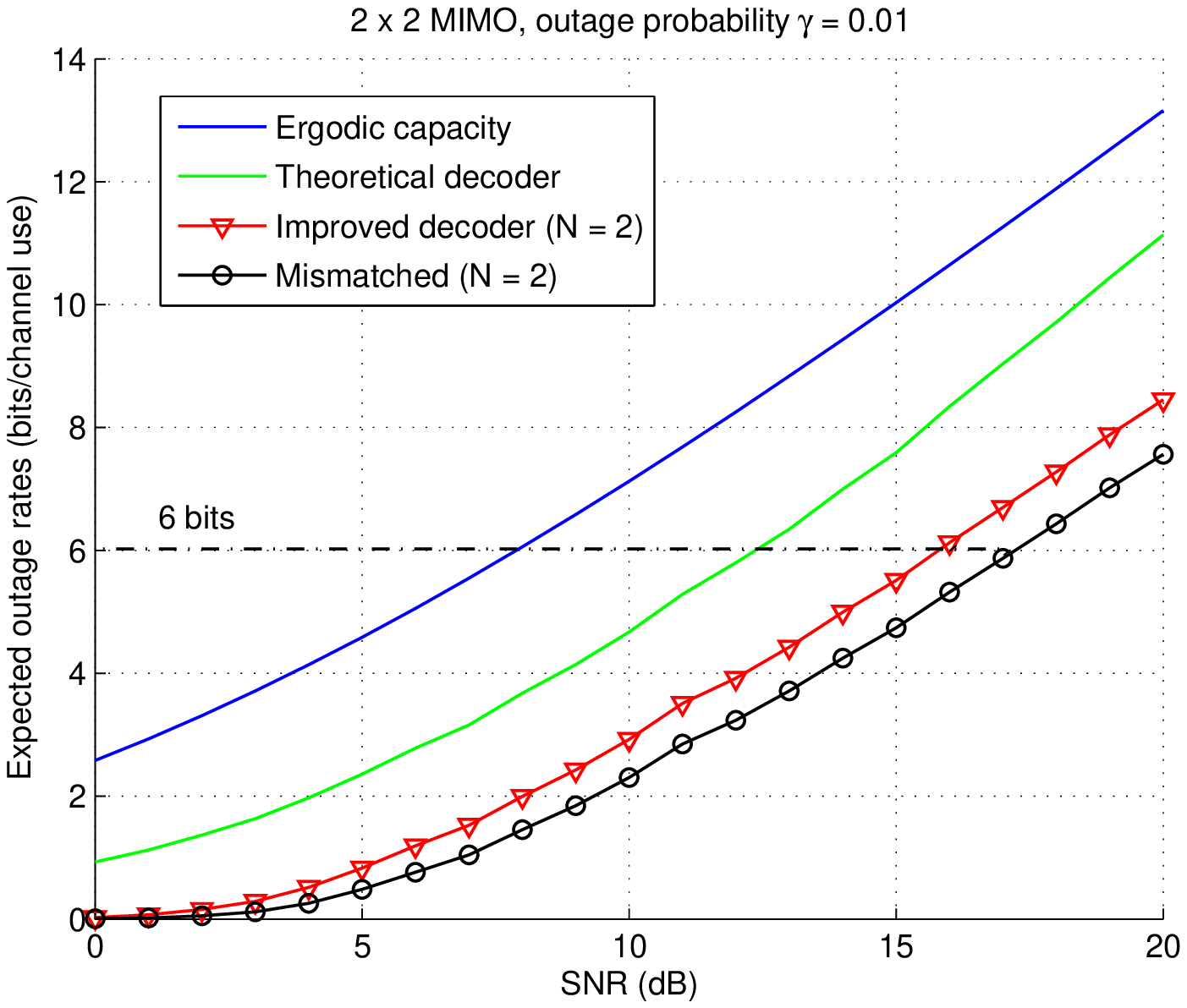}  \vspace{-2mm}
\caption{Expected outage rates of $2\times2$ MIMO system versus SNR $(N=2)$. } \label{cap22}
\end{figure}

Fig. \ref{cap22} compares average outage rates over all channel estimates, of both mismatched ML decoding (given by expression \eqref{acievable_rates_MIMO} and \eqref{mismached_ML_capa}) and our decoding metric (given by \eqref{acievable_rates_MIMO} and \eqref{solution_mu}) versus the SNR. The $2\times2$ MIMO channel is estimated by sending 2 pilot symbols per frame. The outage probability has been fixed to $\gamma_{_{QoS}}=0.01$. For comparison, we also display the upper bounds on these achievable outage rates,  i.e. the EIO capacity (obtained by evaluating \eqref{eq-max}) and the ergodic capacity with perfect channel knowledge at the decoder. It can be observed that the achievable rate using the mismatched ML decoding is about $5$ dB (at a mean outage rate of 6 bits) of SNR far from the EIO capacity. Also, we note that the investigated decoder achieves higher rates for any SNR values and decreases by about $1.5$ dB the aforementioned SNR gap.

\begin{figure}[!htb]
\vspace{-8mm}
\centering
\includegraphics[width=0.5\textwidth,height=0.25\textheight]{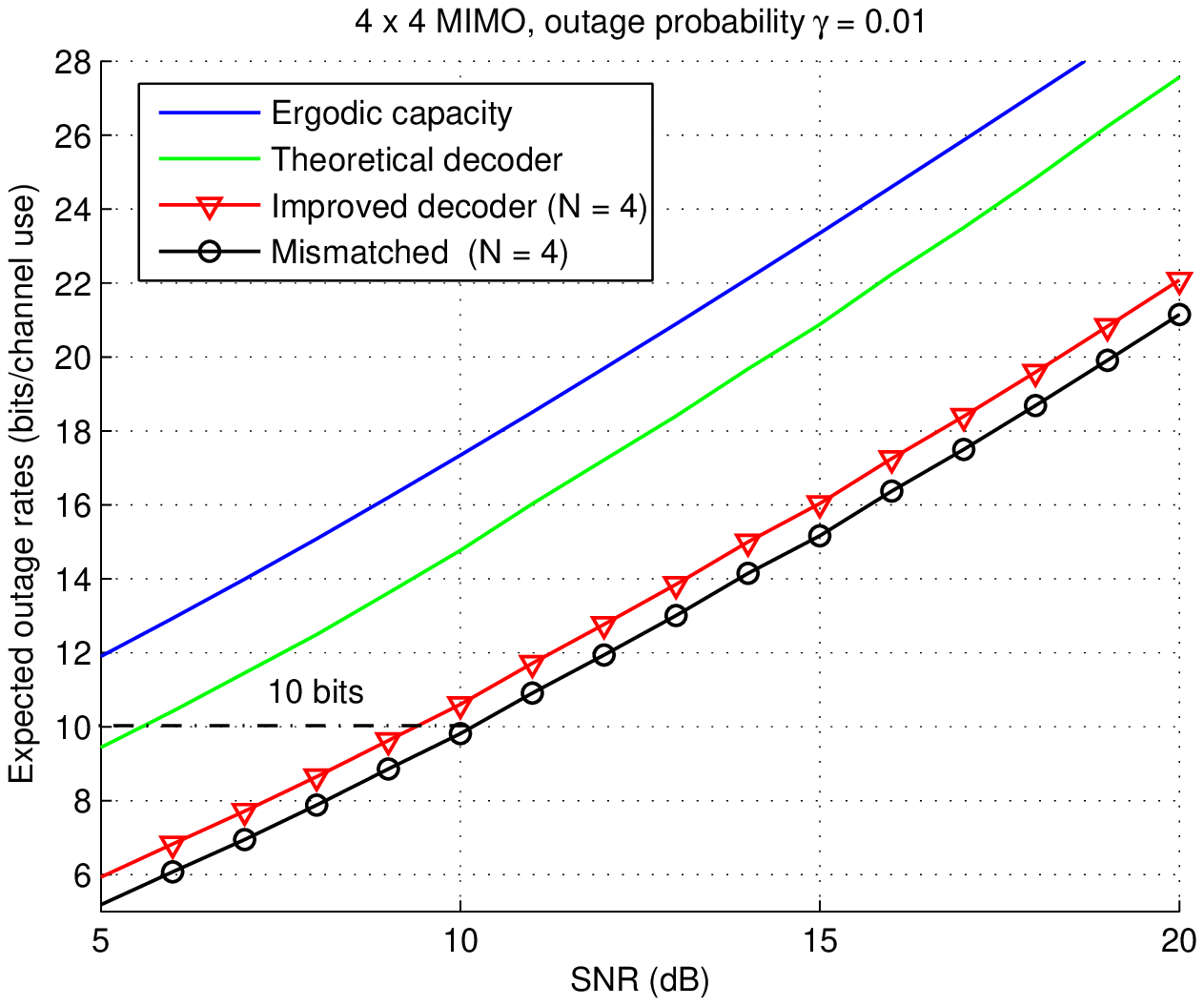}  \vspace{-2mm}
\caption{Expected outage rates of $4\times4$ MIMO system versus SNR $(N=4)$.} \vspace{-8mm}  \label{cap44}
\end{figure}

Similar plots are shown in Fig. \ref{cap44} for a $4\times4$ MIMO channel estimated with $N=4$. Again, it can be observed that the modified decoder achieves higher rates than the mismatched decoder. However, the performance degradation of the mismatched compared to the improved decoder has decreased to less than $1$dB (at $10$ bits). This is a consequence of using orthogonal training sequences with $N\geq M_T$ and the fact that channel estimation is improved by increasing number of antennas \cite{garg05}.\vspace{-3mm}

\section{Conclusion}
\label{sec:concl}
This paper studied the problem of reception in practical communication systems,
when the receiver has only access to a noisy estimate of the channel and this is
not available at the transmitter. By minimizing the average of the transmission error
probability over all channel estimation errors, we derived an improved decoder
adapted to the imperfect channel estimation. Although we showed that the proposed
decoder outperforms the classical mismatched approach, the derivation of a practical
 decoder  achieving the EIO capacity (maximizing over all possible decoders) remains
 as an open problem.

We also derived the expression of the achievable rates associated to the improved
 decoder and compare these to the classical mismatched ML decoding, which replaces
  the perfect channel by its imperfect estimate. As a practical application, the improved
   decoder is used for iterative BICM decoding of MIMO under imperfect channel knowledge.
   Simulation results over Rayleigh block fading MIMO channels indicate that mismatched
    ML decoding is sub-optimal, in terms of BER and achievable rates for short training
     sequences, and confirmed the adequacy of the improved decoder. This performance
     improvement was obtained without introducing any additional complexity. \vspace{-3mm}


\section*{Acknowledgment}
The authors would like to gratefully acknowledge discussions with Walid Hachem.\vspace{-2mm}


%
\bibliographystyle{IEEEbib}
\bibliography{../../bibliography/biblio}

\begin{thebibliography}{1}

\bibitem{isita2006}
P.~Piantanida, G.~Matz, and P.~Duhamel,
\newblock ``Outage behavior of discrete memoryless channels under channel
  estimation errors,''
\newblock in {\em Proc. of Int. Sym. on Information Theory and its
  Applications, ISITA 2006}, Oct. 2006.

\bibitem{Tarokh-1999}
V.~Tarokh, A.~Naguib, N.~Seshadri, and A.R. Calderbank,
\newblock ``Space-time codes for high data rate wireless
  communication:performance criteria in the presence of channel estimation
  errors,mobility, and multiple paths,''
\newblock {\em IEEE Transactions on Communications}, , no. 2, pp. 199--207, Feb
  1999.

\bibitem{taricco-biglieri-2005}
G.~Taricco and E.~Biglieri,
\newblock ``Space-time decoding with imperfect channel estimation,''
\newblock {\em IEEE Trans. on Wireless Communications}, vol. 4, no. 4, pp. 2426
  -- 2467, July 2005.

\bibitem{spawc06}
P.~Piantanida, G.~Matz, and P.~Duhamel,
\newblock ``Estimation-induced outage capacity of ricean channels,''
\newblock in {\em Proc. of Signal Processing for Advanced Wireless
  Communications (SPAWC)}, July 2006.

\bibitem{boutros00}
J.~J. Boutros, F.~Boixadera, and C.~Lamy,
\newblock ``Bit-interleaved coded modulations for multiple-input
  multiple-output channels,''
\newblock in {\em Int. Symp. on Spread Spectrum Tech. and Applications}, Sept.
  2000, pp. 123--126.

\bibitem{bcjr}
L.~Bahl, J.~Cocke, F.~Jelinek, and J.~Raviv,
\newblock ``Optimal decoding of linear codes for minimizing symbol error
  rate,''
\newblock {\em IEEE Trans. Information Theory}, pp. 284--287, March 1974.

\bibitem{merhav-1994}
N.~Merhav, G.~Kaplan, A.~Lapidoth, and S.~Shamai~(Shitz),
\newblock ``On information rates for mismatched decoders,''
\newblock {\em IEEE Trans. Information Theory}, vol. IT-40, no. 6, pp.
  1953--1967, Nov. 1994.

\bibitem{convex-book}
J.B. Hirriart-Urruty and C.~Lemar\'{e}chal,
\newblock {\em Convex Analysis and Minimization Algorithms I},
\newblock Springer-Verlag, 1993.

\bibitem{garg05}
P.~Garg, R.~K. Mallik, and H.~M. Gupta,
\newblock ``Performance analysis of space-time coding with imperfect channel
  estimation,''
\newblock {\em IEEE Trans. Wireless Commun.}, vol. 4, pp. 257--265, Jan. 2005.

\end{thebibliography}

\end{document}